\documentclass[aps,12pt,floatfix,longbibliography,a4paper]{revtex4-2}

\usepackage{graphicx}
\usepackage{epstopdf}
\usepackage{amssymb}
\usepackage{hyperref}
\usepackage[normalem]{ulem}
\usepackage{amsmath,bm}
\usepackage[utf8]{inputenc}
\usepackage{xcolor}
\usepackage{physics}
\usepackage{mathcomp}
\usepackage{multirow}
\usepackage{comment}
\usepackage{makecell}

\usepackage{placeins}

\makeatletter
\@addtoreset{figure}{hoge}
\makeatother
\makeatletter
\@addtoreset{equation}{hoge}
\makeatother
\makeatletter
\@addtoreset{section}{hoge}
\makeatother
\makeatletter
\@addtoreset{table}{hoge}
\makeatother

\begin{document}

\title[]{Observation of \textit{g}-wave altermagnetic multipole}


\author{Ryo Misawa$^{1}$}
\email{misawann6@g.ecc.u-tokyo.ac.jp}
\author{Rikuto Oiwa$^{2}$}
\author{Shunsuke Kitou$^{3}$}
\author{Tatsuya Miki$^{4}$}
\author{Motohiko Ezawa$^{1}$}
\author{Weiyi Yun$^{1}$}
\author{Rinsuke Yamada$^{1}$}
\author{Chihaya Koyama$^{3}$}
\author{J. Alberto Rodríguez Velamazán$^{5}$}
\author{Kamil K. Kolincio$^{6}$}
\author{Navid Qureshi$^{5}$}
\author{Elina Zhakina$^{1}$}
\author{Yuiga Nakamura$^{7}$}
\author{Jan Masell$^{8}$}
\author{Ilya Belopolski$^{9}$}
\author{Taka-hisa Arima$^{3,10}$}
\author{Yusuke Nomura$^{4,11}$}
\author{Satoru Hayami$^{2}$}
\author{Max Hirschberger$^{1,10}$}
\email{hirschberger@ap.t.u-tokyo.ac.jp}

\affiliation{$^{1}$Department of Applied Physics and Quantum-Phase Electronics Center, The University of Tokyo, Bunkyo, Tokyo 113-8656, Japan}
\affiliation{$^{2}$Graduate School of Science, Hokkaido University, Sapporo 060-0810, Japan}
\affiliation{$^{3}$Department of Advanced Materials Science, The University of Tokyo, Kashiwa, Chiba 277-8561, Japan}
\affiliation{$^{4}$Institute for Materials Research (IMR), Tohoku University, Sendai 980-8577, Japan}
\affiliation{$^{5}$Institut Laue-Langevin, 71 Avenue des Martyrs, 38000 Grenoble, France}
\affiliation{$^{6}$Gdańsk University of Technology, Faculty of Applied Physics and Mathematics, Narutowicza 11/12, 80-233 Gdańsk, Poland}
\affiliation{$^{7}$Japan Synchrotron Radiation Research Institute (JASRI), SPring-8, Hyogo 679-5198, Japan}
\affiliation{$^{8}$Institute of Theoretical Solid State Physics, Karlsruhe Institute of Technology (KIT), 76049 Karlsruhe,
Germany}
\affiliation{$^{9}$School of Electrical \& Electronic Engineering, Nanyang Technological University, Singapore}
\affiliation{$^{10}$RIKEN Center for Emergent Matter Science (CEMS), Wako, Saitama 351-0198, Japan}
\affiliation{$^{11}$Advanced Institute for Materials Research (WPI-AIMR), Tohoku University, Sendai 980-8577, Japan}

\maketitle
\newpage

\begin{center}
\Large{Abstract}
\end{center}

\textbf{Over the past few years, altermagnets have emerged as a new class of collinear magnets with broken time-reversal symmetry, offering novel opportunities for spintronics beyond conventional magnets.
Rather than from net magnetization, as in ferromagnets, the unconventional time-reversal symmetry breaking of altermagnets originates from antiferroic magnetic dipoles locked to higher-order multipoles. Here we report the direct visualization of a \textit{g}-wave altermagnetic multipole in the canonical altermagnet CrSb. Combining high-energy synchrotron X-ray diffraction with valence electron density (VED) analysis, we uncover a pronounced directional anisotropy of the VED distribution alternating between Cr sublattices. This evidences the antiferroic order of electric hexadecapoles predicted in \textit{g}-wave altermagnets. Its coexistence with antiferroic magnetic dipoles induces ferroic magnetic multipoles, as probed by polarized neutron diffraction. We further identify a microscopic model of altermagnetism that directly relates the \textit{g}-wave multipole and the \textit{g}-wave spin splitting. Through direct observation and quantification of multipoles, this study provides a real-space fingerprint of altermagnetism and establishes a general probe of hidden multipole order in quantum materials.}

\newpage

\begin{center}
\Large{Main Text}
\end{center}

When an atom is isolated in a vacuum, its valence electron density (VED) distribution is isotropic because atomic orbitals remain degenerate in energy. In a crystal, the orbital degeneracy is lifted by the electric crystal potential of the surrounding atoms, rendering the VED anisotropic. This anisotropy is expressed in terms of multipoles, or equivalently spherical harmonics labeled by the quantum numbers $l$ and $m$~\cite{Hutchings1964-gz,Kusunose2008-lj}. Historically, multipoles have long served as an order parameter for ``hidden" electronic phases in correlated $f$-electron systems, where higher-order multipoles, such as quadrupoles $(l = 2)$ and hexadecapoles $(l = 4)$, can dominate the low-energy physics~\cite{Santini2009-gp,Aeppli2020-vn}. Such hidden orders elude detection by conventional experimental methods and are probed indirectly~(see Supplementary Information).

More recently, spin-group theory has identified altermagnets as a third class of collinear magnets that break time-reversal symmetry despite zero net magnetization~\cite{Smejkal2020,Smejkal2022,Smejkal2022Emerg,Naka2019}, with applications for efficient information processing~\cite{Rafael2021,Smejkal2022TMR}. The multipole language provides a natural description of this phenomenology, reflecting the coupling between electron spins and the local electric crystal potential. In real space, altermagnets carry electric multipoles with $d$-wave $(l = 2)$, $g$-wave $(l = 4)$, or $i$-wave $(l = 6)$ anisotropy~\cite{Verbeek2024,Bhowal2024,McClarty2024,Fernandes2024,Buiarelli2025,Schiff2025,Ubiergo2025,Jungwirth2026-uf}, rather than the isotropic $s$-wave $(l = 0)$ character typical of ferromagnets. The interlocking of this multipole order with electron spins breaks time-reversal symmetry~\cite{Verbeek2024,Jungwirth2026-uf}.

\begin{figure}[ht]%
\centering
\includegraphics[width=0.8\textwidth]{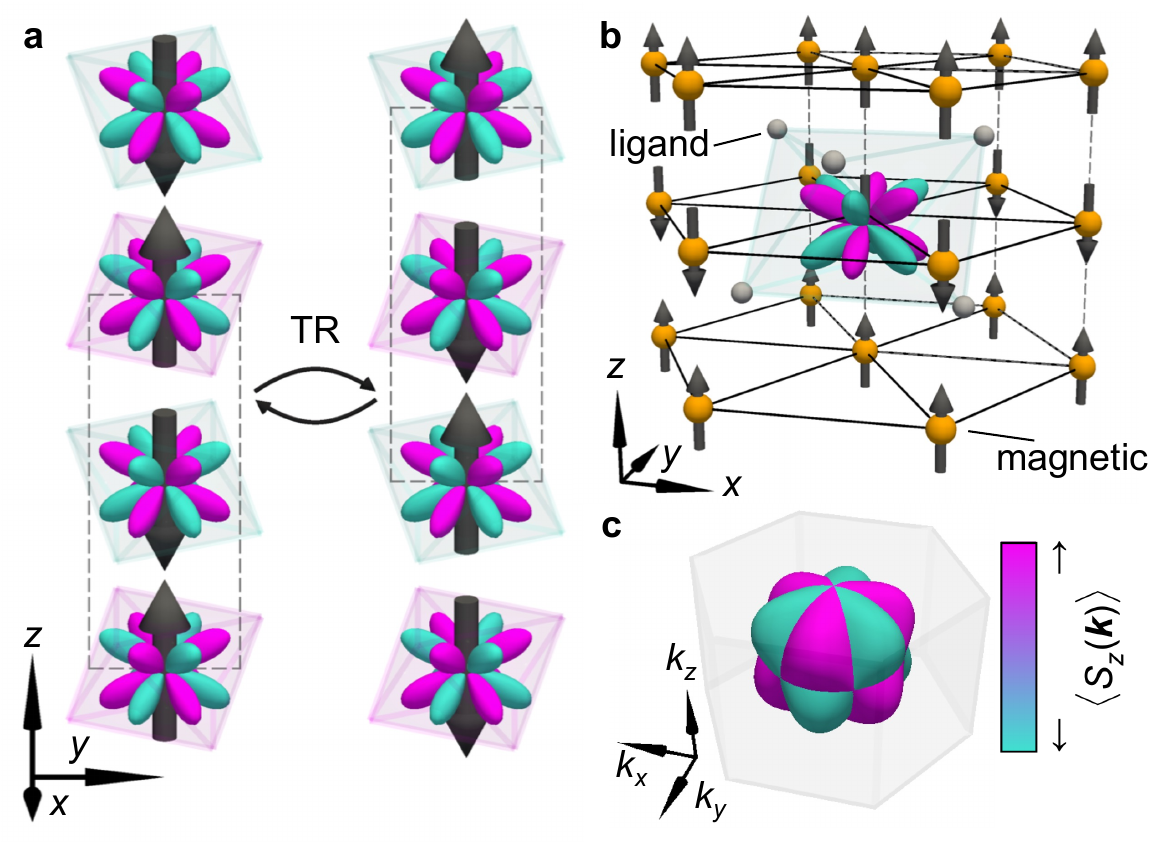}
\caption{\textbf{Multipole description of \textit{g}-wave altermagnetism.}
\textbf{a}, Concept of $g$-wave altermagnetism. The $g$-wave multipoles with $yz(3x^2-y^2)$ anisotropy (positive, magenta; negative, turquoise) and spins (black arrows) are stacked in alternating fashion with opposite sign along the $z$ axis. Due to this antiferroic arrangement of the electric multipoles, the time-reversal (TR) partnered magnetic structures are not equivalent: the spins reverse, but the $g$-wave multipoles do not, under the TR operation. Combined TR and translation symmetry is broken, unlike in a conventional antiferromagnet. The gray dotted line indicates the magnetic unit cell. 
\textbf{b}, Prototypical hexagonal crystal structure of a $g$-wave altermagnet. The antiferroic order of $g$-wave multipoles at magnetic sites, shown on a single site, is predicted for this class of materials. Here, the $g$-wave multipole order is induced by the electric crystal potential from the ligand.
\textbf{c}, Fermi surface of a $g$-wave altermagnet with $k_yk_z(3k_x^2-k_y^2)$ anisotropy, which echoes the symmetry of the $g$-wave multipole in real space. Spin-up and spin-down Bloch states are shown in magenta and turquoise, respectively.
}
\label{main:Fig1}
\end{figure}

To illustrate this central concept of altermagnetism, Fig.~\ref{main:Fig1}\textbf{a} shows an antiferroic pattern of electric hexadecapoles with $yz(3x^2-y^2)$ anisotropy $(l=4)$, predicted in altermagnets with bulk $g$-wave anisotropy~\cite{Verbeek2024,Ubiergo2025,Jungwirth2026-uf} and hereafter referred to as $g$-wave multipoles (see Supplementary Information). These $g$-wave multipoles arise from the crystal potential created by a distorted octahedral environment, as shown in Fig.~\ref{main:Fig1}\textbf{b}. Under the time reversal (TR) operation, magnetic dipoles reverse while the multipoles remain unchanged, rendering the two TR-partnered states inequivalent and breaking TR symmetry. In other words, the concurrent antiferroic order of the electric multipoles and magnetic dipoles gives rise to ferroic magnetic multipoles as an order parameter of altermagnets (see Supplementary Information)~\cite{Verbeek2024}. This differs from a conventional antiferromagnet, which is mapped onto itself by TR with a half-translation. 

As a consequence of the intertwined multipole and magnetic order, altermagnets exhibit anisotropic spin splitting of electronic bands in momentum space~\cite{Verbeek2024,Jungwirth2026-uf}. Remarkably, the splitting can reach the electron-volt scale~\cite{Yang2025}, comparable to that of ferromagnets. In altermagnets with bulk $g$-wave symmetry, the spin splitting follows the form $k_yk_z(3k_x^2-k_y^2)$~(Fig.~\ref{main:Fig1}\textbf{c}), mirroring the anisotropy of the $g$-wave multipole (Fig.~\ref{main:Fig1}\textbf{a}). This momentum-space picture of altermagnets has been studied extensively using angle-resolved photoemission spectroscopy~\cite{Krempasky2024,Ding2024,Reimers2024,Yang2025}. However, the essential real-space hallmark of altermagnets---multipole order reflected in the valence electron and spin density---remains unresolved experimentally.

\begin{figure}[ht]%
\centering
\includegraphics[width=0.7\textwidth]{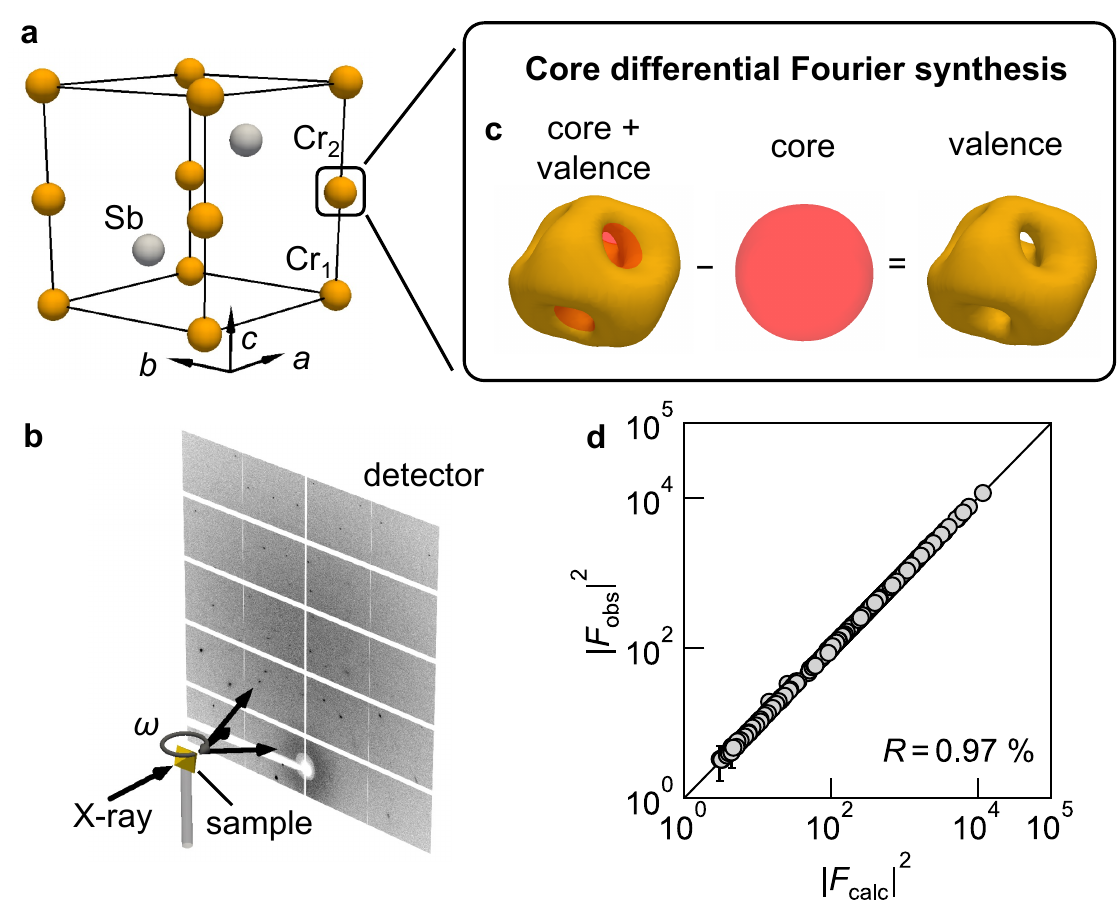}
\caption{\textbf{Real-space visualization of valence electron density (VED) distribution by synchrotron X-ray diffraction.} 
\textbf{a}, Hexagonal $P6_{3}/mmc$ structure of the canonical $g$-wave altermagnet CrSb.
\textbf{b}, Schematic of synchrotron X-ray diffraction, where intensities are recorded with an area detector.
\textbf{c}, Concept of the core differential Fourier synthesis (CDFS) method. A VED distribution is obtained by performing an inverse Fourier transform on crystal structure factors after subtracting a core electron density from the observed total electron density. \textbf{d}, High-quality structural refinement on our CrSb crystals forming a basis for VED analysis. The agreement between the observed and calculated diffraction intensities, expressed as $|F_{\mathrm{obs}}|^{2}$ and $|F_{\mathrm{calc}}|^{2}$, is quantified by a reliability factor $R$ below 1\%. Here, $F$ denotes the structure factor.
}

\label{main:Fig2}
\end{figure}

\bigskip
\textbf{Valence electron density measurement in \textit{g}-wave altermagnet CrSb}\\
Our target material, the canonical room-temperature altermagnet CrSb, crystallizes in the hexagonal $P6_{3}/mmc$ structure with two Cr sublattices of trigonal $\bar{3}m$ site symmetry (Fig.~\ref{main:Fig2}\textbf{a})~\cite{Park2020,Smejkal2022,Smejkal2022Emerg,Ubiergo2025}. Both theoretical and experimental studies have shown giant $g$-wave spin splitting of the band structure~\cite{Ding2024,Reimers2024,Yang2025}. Furthermore, the control of altermagnetic order is demonstrated through crystal-symmetry engineering~\cite{Zhou2025}. As such, CrSb serves as a benchmark platform for revealing multipoles as the real-space fingerprint of altermagnetism, by measuring the VED distribution.

To directly visualize a VED distribution around the Cr sites in CrSb, we employ high-energy synchrotron X-ray diffraction (Fig.~\ref{main:Fig2}\textbf{b}) together with the core differential Fourier synthesis (CDFS) method~\cite{Kitou2017-dk,Kitou2020-vw}. From the total electron density imprinted in the X-ray diffraction data, the CDFS method extracts the VED distribution by subtracting the core electron density, without introducing assumptions about the valence states. Figure~\ref{main:Fig2}\textbf{c} illustrates the principle of the CDFS technique; see Methods for details.
 
Figure~\ref{main:Fig2}\textbf{d} displays our crystal structure refinement for CrSb at $100\,$K. The observed and calculated intensities show exceptionally good agreement with an $R$ factor of $0.97\,\%$. This provides a solid foundation for the subsequent VED analysis using the CDFS method.

\begin{figure}[ht]%
\centering
\includegraphics[width=0.99\textwidth]{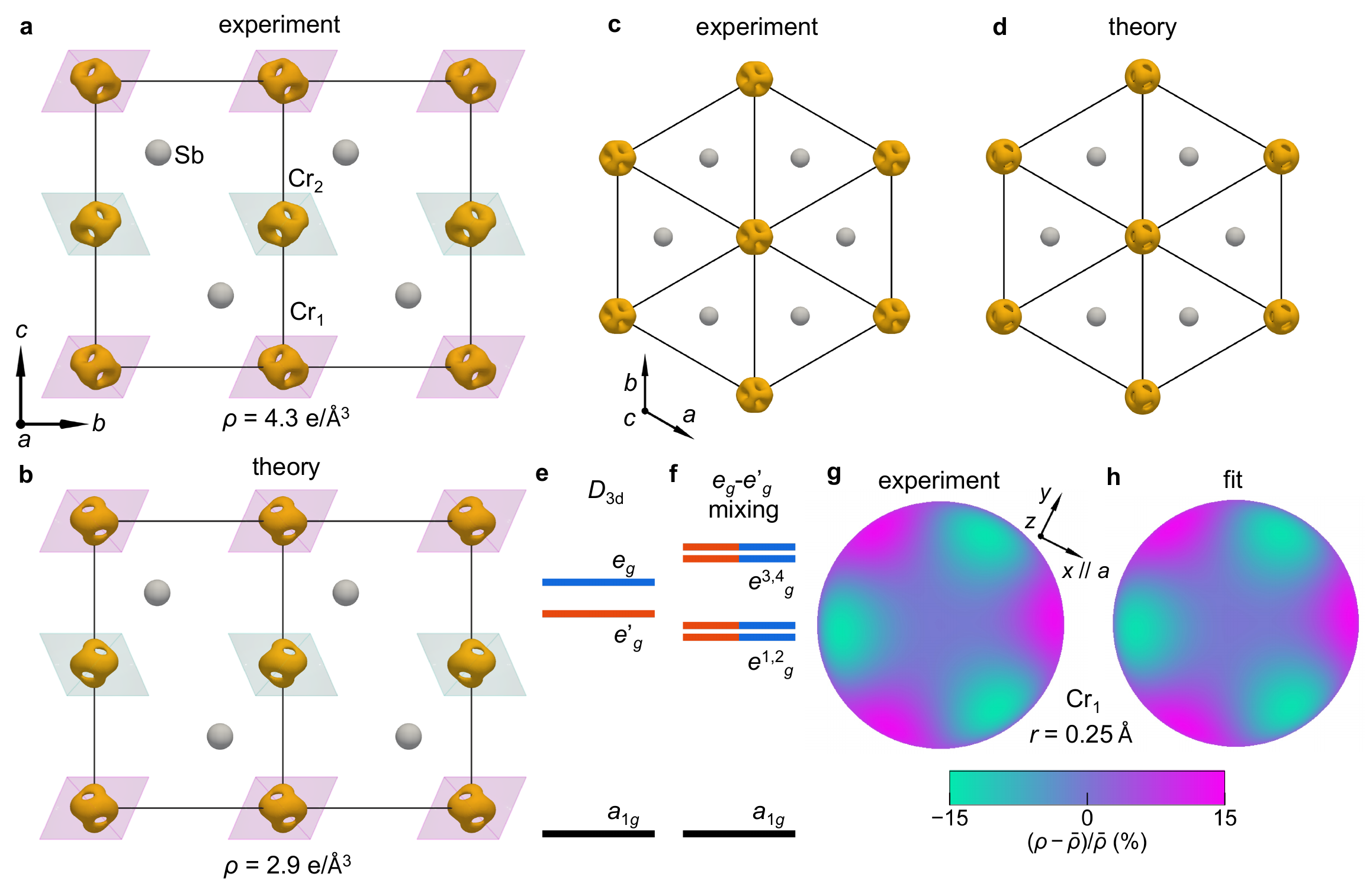}
\caption{\textbf{Direct observation of anisotropic valence electron density (VED) in a \textit{g}-wave altermagnet.}
\textbf{a},\textbf{b}, Side views of the VED distribution at Cr sites from experiment and electronic structure calculations, with isosurface values written below the map (see Methods). In line with theory, the experimental VED shows an alternating arrangement between two Cr sublattices. The magenta and turquoise octahedra highlight the alternating ligand environment of the two Cr sublattices.
\textbf{c},\textbf{d}, Top views of the VED distribution in the Cr$_1$ layer with threefold rotation symmetry, obtained from experiment and theory. For the experimental VED, the isosurface is displayed after an isotropic magnification by a factor of $1.4$ along each axis for visual clarity (see Supplementary Information).
\textbf{e}, Energy-level scheme of Cr $d$ orbitals under trigonal $D_\mathrm{3d}$ point group symmetry. A trigonal distortion of the octahedral environment splits the Cr $d$ manifold into $a_{1g}$, $e_g'$, and $e_g$ levels.
\textbf{f}, Energy-level scheme for eigenstates $e_g^{1\text{-}4}$, arising from the mixing of $e_g$ and $e_g'$ orbitals with similar energies.
\textbf{g},\textbf{h}, Angular dependence of the valence electron density (VED) on a sphere of fixed radius $r_0 = 0.25~\text{\AA}$ centered at the Cr$_1$ nucleus, $\rho(r=r_0,\theta,\phi)$, obtained experimentally and from a fit to the data based on the level scheme shown in panel~\textbf{f}. The line of view is the same as in panel~\textbf{c}, and arrows indicate the orbital coordinate axes. The color scale corresponds to the relative anisotropy of the VED, $\left[\rho(r_0, \theta, \phi)-\overline{\rho(r_0, \theta, \phi)}\right]/\overline{\rho(r_0, \theta, \phi)}\times 100$, where $\overline{\rho(r_0, \theta, \phi)}$ is the angle-averaged VED.
}
\label{main:Fig3}
\end{figure}

\bigskip
\textbf{Observation of \textit{g}-wave altermagnetic VED}\\
Figure~\ref{main:Fig3}\textbf{a} shows the iso-density surface of the experimentally observed VED around the two Cr sublattices, corresponding to valence electrons occupying Cr $3d$ orbitals. Focusing on a single Cr site, low values of the VED are located along the directions towards the surrounding six Sb atoms. These six holes in the isosurface are consistent with a local ionic picture, where the Cr-centered VED is depleted along the ligand directions, rather than with a strongly covalent picture that would produce appreciable interstitial VED along the Cr-Sb bonds.

Seen from the side, the VED exhibits an alternating arrangement along the $c$ axis. For comparison, we perform electronic-structure calculations, and the resulting VED distribution reproduces the characteristic anisotropy in both the side (Fig.~\ref{main:Fig3}\textbf{b}) and top views (Fig.~\ref{main:Fig3}\textbf{c},\textbf{d}). The VED on the Sb sites cannot be reconstructed due to strongly delocalized $5p$ orbitals; see Methods for details.

To understand the microscopic origin of the anisotropic VED, we next consider an energy-level scheme of Cr $d$ orbitals. In a trigonally distorted octahedral environment ($D_\mathrm{3d}$) formed by the Sb ligand, the electric crystal potential splits energy levels into $e_g$ ($d_{x^2-y^2}$, $d_{xy}$), $e_g'$ ($d_{yz}$, $d_{zx}$), and $a_{1g}$ ($d_{3z^2-r^2}$) states (Fig.~\ref{main:Fig3}\textbf{e}). Furthermore, our band-structure calculations yield a small bare separation between $e_g$ and $e_g'$ ($0.015~$eV). This leads to mixing of the $e_g$ and $e_g'$ manifolds into two doubly degenerate manifolds, $e_g^{1,2}$ and $e_g^{3,4}$ (Fig.~\ref{main:Fig3}\textbf{f}). The lower-energy states $e_g^{1,2}$ are given by
\begin{align}
\ket{e^1_{g}} &= \alpha c_1\ket{d_{x^2-y^2}} + c_2\ket{d_{yz}}, \\
\ket{e^2_{g}} &= \alpha c_2\ket{d_{zx}} + c_1\ket{d_{xy}},
\end{align}
where $c_1$ and $c_2$ describe the strength of $e_g$-$e_g'$ mixing, $c_1^2 + c_2^2 = 1$, and $\alpha$ denotes the sublattice degree of freedom  ($\alpha=\pm1$ for Cr$_1$, Cr$_2$). The higher-energy state $e_g^{3}$ ($e_g^{4}$) is constructed from $e_g^{1}$ ($e_g^{2}$) by swapping the two coefficients and changing the sign of one, to form an orthogonal basis; see Supplementary Information.

Using this model, we fit the angular dependence of the experimental VED $\rho_\mathrm{exp}(r_0, \theta, \phi)$ at a fixed distance $r_0=0.25~\text{\AA}$ from the Cr nucleus, as shown in Fig.~\ref{main:Fig3}\textbf{g}. 
The position $r_0$ corresponds to the VED maximum in the radial dependence of Cr $3d$ orbitals. In the ionic limit, the formal valence of Cr is $3+$, and there is one valence electron in each of the three orbitals $a_{1g}$ and $e_g^{1,2}$. While this ionic model qualitatively captures the directional dependence of the VED, its anisotropy is much stronger than that observed experimentally (see Supplementary Information). We thus assume partial occupancies of all the orbitals with total electron fillings $n_{a}$, $n_{e,1}$, and $n_{e,2}$ for the $a_{1g}$, $(e_g^{1},e_g^{2})$, and $(e_g^{3},e_g^{4})$ manifolds, respectively. We fit five parameters—the global scale factor $s$, $c_1$, and three electron occupancies—, compute $\rho_{\mathrm{cal}}(r_0,\theta,\phi)$ from the model, and minimize a figure of merit (Methods). This successfully reproduces the experimental anisotropy, including its magnitude, as shown in Fig.~\ref{main:Fig3}\textbf{h}. As a correction to the local ionic picture, electron hopping in metallic CrSb renormalizes these parameters (see Supplementary Information): $(c_1, n_{a}, n_{e,1}, n_{e,2}) = (0.84, 0.7, 1.5, 1.1)$ instead of $(2/3, 1, 2, 0)$.

Importantly, the $g$-wave multipole ($l=4$) arises from the interference of $e_g$ and $e_g'$ orbitals; otherwise, the VED would exhibit the trivial anisotropy associated with $d$ orbitals ($l=2$). Explicitly, the interference terms in $(e^1_{g})^2$ and $(e^2_{g})^2$ produce a part of the VED that is proportional to 
\begin{equation}
\label{eq:hexdecapole}
     \alpha\left\{yz\cdot(x^2-y^2) + 2zx\cdot xy\right\} = \alpha yz(3x^2-y^2), 
\end{equation}
where the factor of $2$ comes from normalization constants of $d$ orbitals (see Supplementary Information). Equation~\eqref{eq:hexdecapole} changes its sign depending on sublattice $\alpha$, corresponding to antiferroic ordering of the $g$-wave multipole.

\begin{figure}[ht]%
\centering
\includegraphics[width=0.9\textwidth]{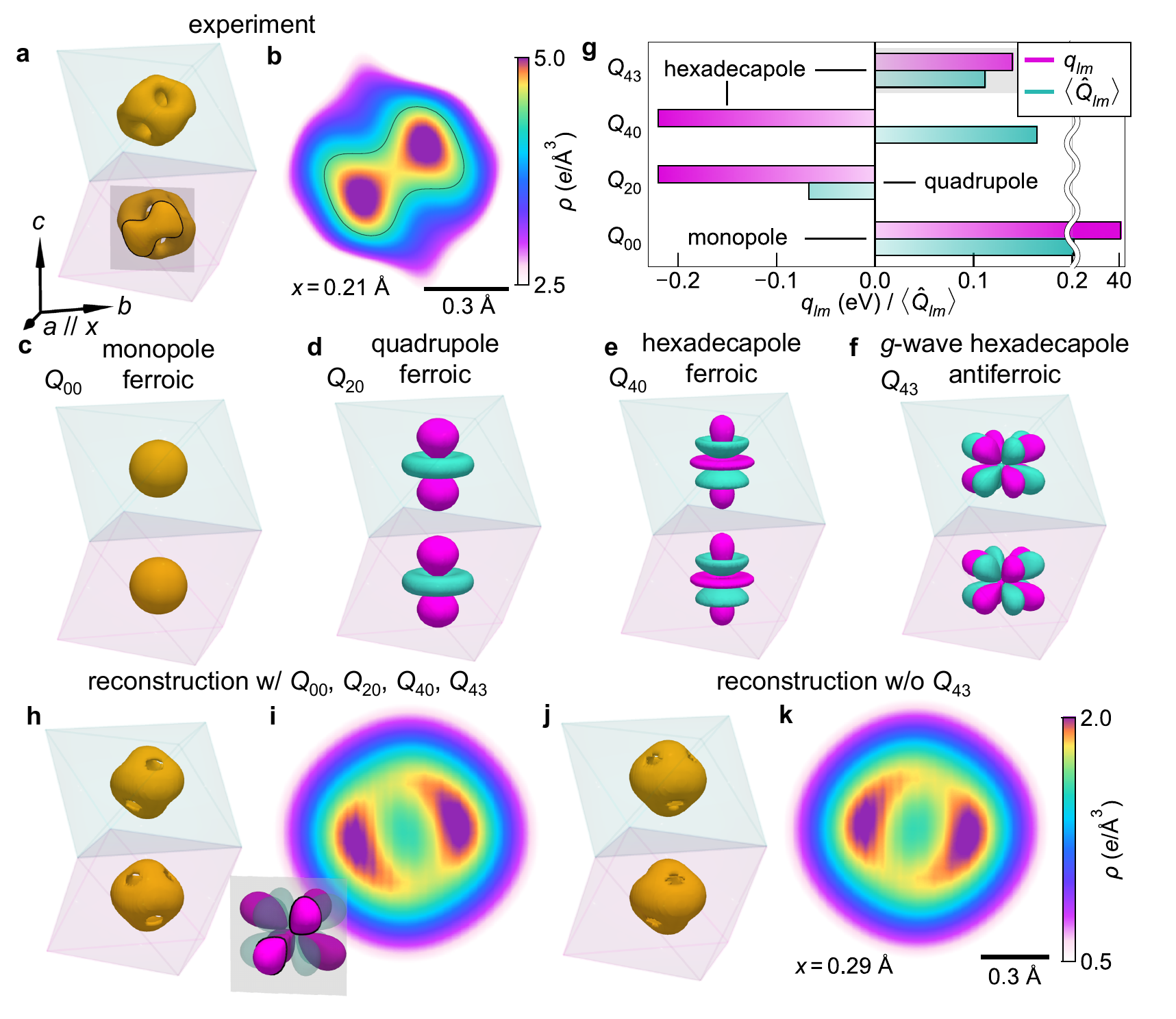}
\caption{\textbf{Multipole decomposition of \textit{g}-wave altermagnetic VED.} \textbf{a},\textbf{b}, Experimental VED at the Cr sites, shown as a yellow isosurface and as a cut on the gray plane. The contour line in panel~\textbf{b} is drawn at the isosurface value in panel~\textbf{a}.
\textbf{c}-\textbf{f}, Active electric multipoles $Q_{lm}$ in CrSb. Here, $Q_{lm}$ denotes the real (tesseral) spherical harmonics with the quantum numbers $l$ and $m$. Only the hexadecapole $Q_{43}$ can show antiferroic order. The isosurfaces are drawn using the radial part of $Q_{lm}$ for an isolated Cr atom (Methods).
\textbf{g}, Values of $q_{lm}$ and $\langle \hat{Q}_{lm}\rangle$ from multipole decomposition based on the symmetry-adapted Wannier formalism (Methods). Here, $\hat{Q}_{lm}$ is the electric crystal potential associated with $Q_{lm}$, and $q_{lm}$ is its coefficient [Eq.~\eqref{eq:CEF_main}]. Expectation values $\langle \hat{Q}_{lm}\rangle$ contribute to the VED (Methods). Contributions from Cr $d$ orbitals are shown.
\textbf{h},\textbf{i}, Reconstructed VED distribution including all $Q_{lm}$. In panel~\textbf{i}, the two VED maxima coincide with the positive lobes of $Q_{43}$ (inset), whereas the negative lobes (transparent turquoise in inset) are not visible because the VED is always positive. The cut is taken at a position $1.4$ times farther from the nucleus than in panel~\textbf{b}; see Supplementary Information.
\textbf{j},\textbf{k}, Reconstructed VED excluding $Q_{43}$. The experimental map shows low-VED holes only along the ligand directions, whereas omitting $Q_{43}$ produces holes along incorrect directions. The deviation from an ideal ferroic VED pattern expected in the local limit reflects hybridization with Sb $p$ orbitals in the Wannier orbitals. Even with this hybridization included for a more realistic description of metallic CrSb, the two VED maxima in panel~\textbf{k} do not match those in panel~\textbf{b}, showing the dominant role of $Q_{43}$.
}
\label{main:Fig4}
\end{figure}

\bigskip
\textbf{Evidence for \textit{g}-wave multipole order by multipole decomposition of VED}\\
To further clarify the $g$-wave multipole order in CrSb, we decompose the anisotropic VED into a linear combination of electric multipoles. The experimental VED shows an alternating texture between two sublattices (Fig.~\ref{main:Fig4}\textbf{a},\textbf{b}), suggesting an antiferroic arrangement of electric multipoles. Our symmetry analysis shows that four electric multipoles are active in the Cr $d$-orbital manifold of the paramagnetic state without spin-orbit coupling: the monopole $Q_{00}$ (Fig.~\ref{main:Fig4}\textbf{c}), the quadrupole $Q_{20}$ (Fig.~\ref{main:Fig4}\textbf{d}), and two hexadecapoles, $Q_{40}$ (Fig.~\ref{main:Fig4}\textbf{e}) and $Q_{43}$ (Fig.~\ref{main:Fig4}\textbf{f}). Here, $Q_{lm}$ denotes the real (tesseral) spherical harmonic labeled by the quantum numbers $l$ and $m$; see Supplementary Information for explicit expressions. Importantly, among these electric multipoles, only the hexadecapole $Q_{43} \propto yz(3x^2-y^2)$, termed here $g$-wave multipole, can exhibit antiferroic order; the other active multipoles are restricted to ferroic arrangements. This is because the two Cr sublattices are related by a mirror operation perpendicular to the $c$ axis, under which only $Q_{43}$ changes its sign.

To quantify contributions from $Q_{lm}$ to the VED, we perform symmetry-adapted closest Wannier modeling (SCW)~\cite{Ozaki2024-oz,Oiwa2025-fz}. This method constructs a tight-binding Hamiltonian as a linear combination of complete multipole basis sets based on band-structure calculations; see Methods. Here, we focus on the Hamiltonian for the crystal potential $\mathcal{H}(\alpha)$ in the $d$-orbital space at the Cr sublattices ($\alpha = \pm 1$), which is written as a linear combination of the electric crystal potentials $\hat{Q}_{lm}(\alpha)$,
\begin{equation}
\label{eq:CEF_main}
    \mathcal{H}(\alpha) =
    q_{00}\hat{Q}_{00} +
    q_{20}\hat{Q}_{20} +
    q_{40}\hat{Q}_{40} +
    q_{43}\hat{Q}_{43}(\alpha),
\end{equation}
where the matrix form of $\hat{Q}_{lm}$ is provided in Supplementary Information. 
We then calculate the coefficients $q_{lm}$ and the expectation values $\langle \hat{Q}_{lm}\rangle$ by SCW modeling. Figure~\ref{main:Fig4}\textbf{g} shows the results of the SCW analysis. Remarkably, the magnitude of the $g$-wave multipole $Q_{43}$ is comparable to $Q_{40}$ and even to the lower-order multipole $Q_{20}$. Subsequently, we reconstruct the VED distributions using a linear combination of $\hat{Q}_{lm}(\alpha)$ (Methods). Including all the multipoles $Q_{lm}$ reproduces the alternating VED pattern (Fig.~\ref{main:Fig4}\textbf{h}). Taking its two-dimensional cut, the two VED maxima coincide with the two positive lobes of $Q_{43}$ (Fig.~\ref{main:Fig4}\textbf{i}), in line with the experimental map (Fig.~\ref{main:Fig4}\textbf{b}). By contrast, removing $Q_{43}$ makes the reconstruction inconsistent with experiment: the low-VED holes no longer align with the Sb directions (Fig.~\ref{main:Fig4}\textbf{j}), and the VED maxima are incorrectly reproduced in the cut (Fig.~\ref{main:Fig4}\textbf{k}). 

\bigskip
\textbf{Observation of ferroic magnetic multipoles}\\
Having demonstrated the antiferroic electric multipoles in CrSb, we show that their coexistence with antiferroic magnetic dipoles produces ferroic magnetic multipoles as an order parameter of the altermagnet. To this end, we perform polarized neutron diffraction measurements of the SD distribution $\rho_\mathrm{s}$ at $300\,$K and a conventional multipole refinement (Methods). We measure the flipping ratio ($\mathrm{FR}$), defined as the ratio of spin-up to spin-down neutron intensities, to detect weak signals associated with magnetic multipoles. The dipole part of a magnetic multipole describes the spin direction, or the sign of $\rho_\mathrm{s}$, while $Q_{lm}$ describes its angular dependence. As opposed to the multipole decomposition of the VED, $Q_{00}$, $Q_{20}$, and $Q_{40}$ components are antiferroic while $Q_{43}$ is ferroic for the SD. This difference arises because magnetic multipoles on two sublattices are related by a combined mirror and time-reversal operation: the additional sign change under time reversal swaps the ferroic and antiferroic arrangements. Due to these symmetry constraints, $Q_{00}$, $Q_{20}$, and $Q_{40}$ contribute to magnetic structure factors at Miller index $l=\,$odd, while they do not affect $l=\,$even. Thus, a deviation of $\mathrm{FR}$ from $1$ at $l=\,$even evidences the ferroic order of magnetic multipoles.

\begin{figure}[ht]%
\centering
\includegraphics[width=0.9\textwidth]{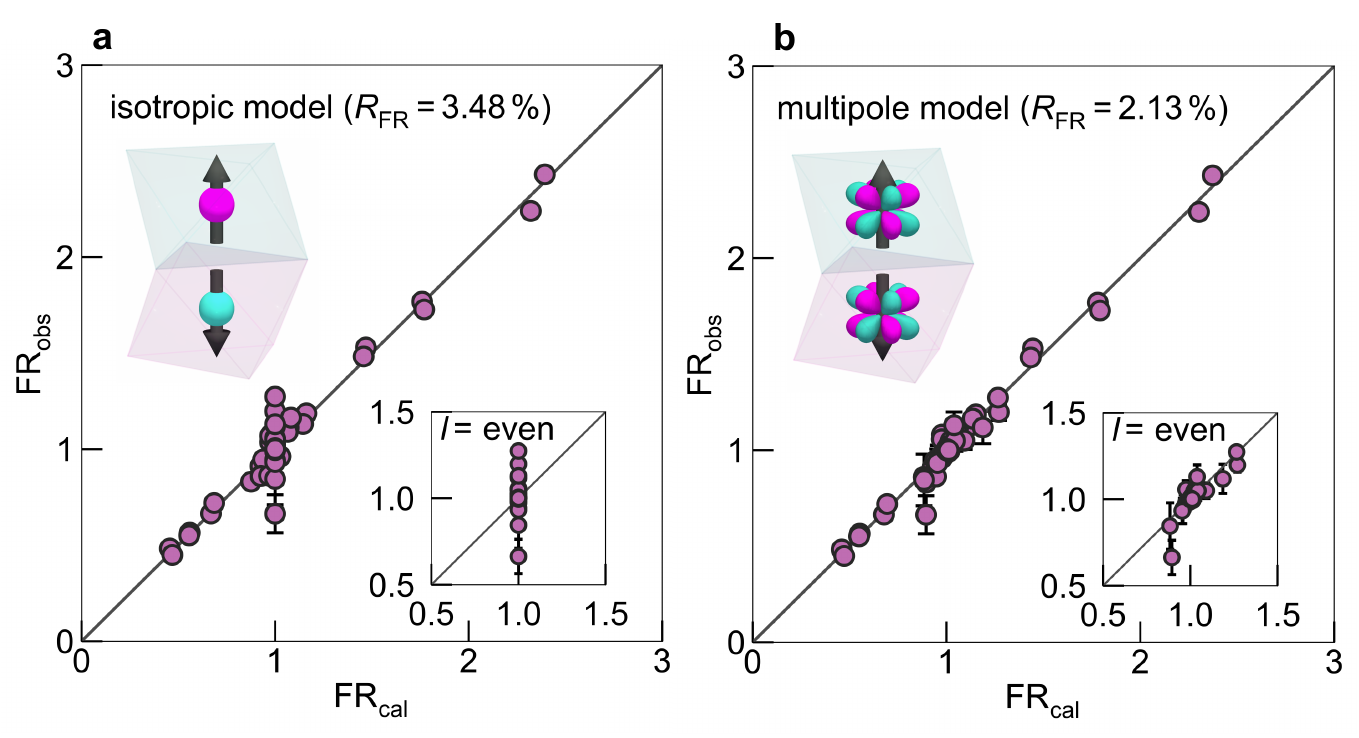}
\caption{\textbf{Ferroic order of magnetic multipoles probed by polarized neutron diffraction.} 
\textbf{a}, Multipole refinement of flipping ratios (FR) at $300\,$K assuming an isotropic model for the spin density at the Cr site. The antiferroic dipole moments affect the flipping ratio only at $l=\,$odd. The inset shows $l=\,$even reflections, for which the isotropic model predicts $\text{FR}_\mathrm{cal}=1$, while the data show significant deviations from $1$. \textbf{b}, Multipole refinement when the ferroic order of magnetic multipoles, characteristic of altermagnetism, is incorporated. The model shows better agreement with the data for $l=\,$even reflections (inset). The angular dependence of the magnetic multipole is determined by the corresponding one-rank-lower electric multipole, $Q_{43}$ (inset). Since the spin density changes sign under time reversal, the $Q_{43}$ component is antiferroic in the VED (Figs.~\ref{main:Fig1}\textbf{a},~\ref{main:Fig4}\textbf{f}) but ferroic in the spin density. Coefficients for the other active multipoles, $Q_{20}$ and $Q_{40}$, are also refined. 
}
\label{main:Fig5}
\end{figure}

We illustrate the fitting result of the isotropic model ($Q_{00}$) in Fig.~\ref{main:Fig5}\textbf{a}, corresponding to a trivial antiferromagnetic state. The inset focuses on $l=\,$even reflections; the experimental data exhibit statistically significant deviations from $1$, while the model predicts the flipping ratio $\mathrm{FR}=1$. This observation provides strong evidence for the ferroic order of higher-order magnetic multipoles (triakontadipoles)~\cite{Verbeek2024}. The isotropic model results in $R_{\mathrm{FR}}=3.48\,\%$, where $R_{\mathrm{FR}}$ measures the level of agreement between the model and the data. Subsequently, we refine all the active multipole coefficients, leading to good agreement ($R_{\mathrm{FR}}=2.13\,\%$), particularly for $l=\mathrm{even}$ due to the inclusion of $Q_{43}$ (Fig.~\ref{main:Fig5}\textbf{b}). The magnitude of $Q_{43}$ is refined to be $0.21(6)$, in good agreement with $0.25$ from the theoretical SD decomposed into multipoles in the same manner (Methods). Because of the direct link between electric and magnetic multipoles, the anisotropies of the VED and the SD are closely related not only in CrSb but also in other altermagnets based on $3d$ transition metals, as confirmed in theoretical calculations and reconstructed SD distributions  (see Supplementary Information).

\bigskip
\textbf{Anisotropic spin splitting from \textit{g}-wave multipole order}\\
Finally, we demonstrate the connection of the $g$-wave multipole order in real space with the spin splitting of electronic bands. From our SCW multipole analysis, we estimate the magnitude of $q_{43}$ to be $0.14\,\mathrm{eV}$ (Fig.~\ref{main:Fig4}\textbf{g}), comparable to the Cr-Sb nearest-neighbor hopping amplitude ($t'=0.18\,\mathrm{eV}$). In conventional hopping-based models of altermagnetism, this hopping amplitude is the key ingredient for spin splitting~\cite{Naka2019,Hayami2020-hr,Smejkal2022,Roig2024}. In such models, the $g$-wave symmetry originates from anisotropic hopping that induces intersite $e_g$-$e_g'$ mixing (see Supplementary Information). On the other hand, the atomic multipole $Q_{43}$ arises from $e_g$-$e_g'$ mixing on the same atomic site. We identify that the interplay between $Q_{43}$ and inter-sublattice hopping $t$ likewise generates the $g$-wave spin splitting, even when the hopping itself does not carry the $g$-wave anisotropy (see Supplementary Information). Explicitly, the lowest-order contribution to the spin splitting $g(\bm{k})$ is shown to be
\begin{equation}
    g(\bm{k}) \propto J_zq_{43} t^2 k_zk_y(3k_x^2-k_y^2),
\end{equation}
where $J_z$ is the exchange interaction. The spin splitting is therefore proportional to the amplitude of the electric crystal potential associated with the $g$-wave multipole.

\bigskip
\textbf{Discussion}\\
Our imaging approach quantifies orbital hybridization and the relative magnitudes of induced multipoles—quantum parameters that are symmetry-allowed yet whose strength is not fixed a priori. Since spin splitting emerges directly from the underlying multipole order in our microscopic model, measuring the magnitude of multipoles helps us clarify the origin of the large splitting observed in real altermagnets~\cite{Krempasky2024,Ding2024,Reimers2024,Yang2025}. More broadly, the methodology developed here---integrating X-ray diffraction measurements of multipoles with Wannier modeling, complemented by neutron-scattering experiments---readily extends to $d$-wave~\cite{Jiang2025-do,Zhang2025-nz} and supercell altermagnets~\cite{Jaeschke-Ubiergo2024-ci}, as well as to odd-parity magnets~\cite{Hellenes2023-bc,Jungwirth2024-cr} with inversion-breaking multipoles. By demonstrating the direct visualization and quantification of multipoles, our work establishes a general strategy for uncovering hidden multipole order across quantum materials, including long-standing challenges for $f$-electron systems~\cite{Santini2009-gp,Kusunose2011-es,Aeppli2020-vn,Tazai2019-hb}.

\newpage
\bibliography{sn-bibliography}

\newpage

\clearpage

\begin{center}
\Large{Main Text Figures}
\end{center}
\FloatBarrier
\vspace{5mm}

\clearpage

\clearpage

\clearpage

\clearpage

\clearpage

\clearpage
\newpage

\begin{center}
\Large{Methods}
\end{center}

\textbf{Crystal growth and characterization}\\
Single crystals of CrSb were synthesized by the chemical vapor transport (CVT) method with iodine as a transport agent. The elements were mixed in a molar ratio of Cr:~Sb:~I$_2 = 1:~1:~0.1$ and sealed in an evacuated ampoule. The sealed ampoule was then placed in a three-zone furnace, where the low-temperature, medium-temperature, and high-temperature zones were maintained at $T_\mathrm{L}\ $= $800\,^\circ\mathrm{C}$, $T_\mathrm{M}\ $= $825\,^\circ\mathrm{C}$ and $T_\mathrm{H}\ $= $850\,^\circ\mathrm{C}$, respectively. Before reaching the target temperature, the hot and cold ends were replaced and held for $5\,$hours to promote the localization of transport agent I$_2$ at the high-temperature end. The furnace was then set to the target temperatures and maintained for $465\,$hours. Thus, irregularly shaped CrSb single crystals with (00$l$) lateral facets were obtained.

\bigskip
\textbf{Synchrotron X-ray diffraction}\\
Synchrotron XRD measurements were carried out at beamline BL$02$B$1$ of the SPring-$8$ synchrotron radiation facility (Japan). Single crystals of CrSb with a typical size of $\sim 50~\mu$m were used for the measurements. A N$_2$-gas-blowing device was used to cool the crystal to $100\,$K; below $100\,$K, a He-gas blower was employed. Bragg reflections corresponding to interplanar spacings of $d > 0.29~\text{\AA}$ were recorded using a CdTe PILATUS area detector and the \textsc{CrysAlisPro} software~\cite{Agilent-Technologies-Ltd2014-hl} with the fine-slice acquisition method, wherein the reciprocal space was scanned in steps of $\Delta \omega = 0.01^\circ$. Symmetry-equivalent reflections were averaged, and subsequent structure refinements were performed with \textsc{Jana2006}~\cite{Petricek2014-pc}. The crystal structure refinement of CrSb was performed with only high-angle reflections satisfying $\sin \theta / \lambda > 0.6\,\text{\AA}^{-1}$. Since the contribution of delocalized valence electrons to X-ray scattering is negligible in this range~\cite{Kitou2017-dk}, highly reliable displacement parameters were obtained, with uncertainties as small as $10^{-4}\,$\AA$^2$ (see Supplementary Information). These parameters were then used consistently to calculate the total and core electron densities, from which the valence electron density was obtained by the subsequent core differential Fourier synthesis method. The high-angle analysis yields a high-quality refinement, with $R_1 = 0.98\,\%$ and $\mathrm{GOF}=1.66$, consistent with a reliable crystallographic model according to the standard checkCIF/PLATON validation criteria. Allowing the Cr occupancy to relax did not improve the refinement, and the refined occupancy converged to unity within error, indicating a stoichiometric single crystal.

\bigskip
\textbf{Analysis of valence electron density (VED)}\\
The valence electron distribution (VED) in CrSb was reconstructed using the core differential Fourier synthesis (CDFS) method~\cite{Kitou2017-dk,Kitou2020-vw}. This method extracts VED distribution from XRD data, which includes total electron density information, by applying the inverse Fourier transform after subtracting core-electron contributions. Specifically, the procedure involves
\begin{equation}
    \rho(\bm{r}) = \frac{1}{V}\sum_{\bm{G}}\left(|F_\mathrm{obs}(\bm{G})|P - |F_\mathrm{cal.}^\mathrm{core}(\bm{G})|P^\mathrm{core}\right)e^{i\bm{G}\cdot\bm{r}} + \frac{n}{V},
\end{equation}
where $V$ and $n$ denote the unit-cell volume and the total number of valence electrons per unit cell, respectively. Here, $|F_\mathrm{obs}(\bm{G})|$ is the absolute value of the entire structure factor obtained from the experiment (intensity $I\propto |F_\mathrm{obs}(\bm{G})|^2$), $|F_\mathrm{cal.}^\mathrm{core}(\bm{G})|$ is the core contribution, and $P$ ($P^\mathrm{core}$) is the phase factor for the (core) structure factor. Since the phase information is lost in XRD experiments, $P$ and $P_\mathrm{core}$ are calculated as $P = F_\mathrm{cal}(\bm{G})/|F_\mathrm{cal}(\bm{G})|$ and $P^\mathrm{core} = F^\mathrm{core}_\mathrm{cal}(\bm{G})/|F^\mathrm{core}_\mathrm{cal}(\bm{G})|$, respectively, using the calculated structure factors from the structure refinement. Core electrons contribute to a wide range of momentum space, so that the subtraction of their contributions mitigates truncation errors when obtaining the electron density. This method, hence, enables precise reconstruction of VED in crystalline solids.

In these calculations, [Ar]- and [Kr]-type electron configurations were treated as core electrons for Cr and Sb atoms, respectively. Using the atomic form factors for the core electrons and the structural parameters fixed by the high-angle analysis, the core electron density was calculated. The effects of atomic thermal motion were suppressed by performing the measurements at $100\,$K. Indeed, analytical calculations and temperature-dependent XRD measurements show that harmonic atomic thermal displacements cannot generate the $Q_{43}$ anisotropy by symmetry, and that even the leading-order anharmonic contribution is negligible (see Supplementary Information). In the CDFS method, we assume an isotropic core electron density when subtracting its contribution from a structure factor. The validity of this assumption is confirmed by theoretical calculations of the core electron density in CrSb (see Supplementary Information). The three-dimensional voxel size for the reconstructed VED map was set to $\Delta V = (0.02\,\text{\AA})^3$. Owing to the use of high-energy synchrotron X-ray ($38\,\mathrm{keV}$), we achieve a spatial resolution of approximately $0.29\,\text{\AA}$. Since the Sb $5p$ orbitals are strongly delocalized, resolving their VED distribution would require a dynamic range larger than that of the present detector ($\sim 10^6$). As a result, the VED on the Sb sites could not be reliably reconstructed.

To confirm the reproducibility of the altermagnetic VED, we performed the same measurement on another crystal piece. We achieve a comparable $R$-value (see Supplementary Information) and observe a similar VED distribution that alternates at two Cr sublattices (see Supplementary Information).

\bigskip

\textbf{Fitting of the valence electron density (VED)}\\
To fit the experimental VED, we consider the functional form
\begin{equation}
\label{eq:ved}
\rho_{\mathrm{calc}}(r, \theta, \phi) = |R(r)|^2\varphi(\theta,\phi),
\end{equation}
where $R(r)$ and $\varphi(\theta,\phi)$ describe the radial wavefunction and the squared angular wavefunction of the VED, respectively. 
Fixing $r=r_0=0.25\, $\AA, we use $R(r_0)$ from the Slater-type orbital (STO) for the $3d$-orbital wavefunction of an isolated Cr atom~\cite{Su1997-jt}.
With electron fillings $n_{a}, n_{e,1}\,$ and $n_{e,2}$ of $a_{1g}$, $e^{1,2}_{g}$, and $e^{3,4}_{g}$, $\varphi(\theta,\phi)$ is calculated as
\begin{equation}
\label{eq:ved_model}
\varphi(\theta,\phi) = n_{a}|a_{1g}|^2 + \frac{n_{e,1}}{2}\left(|e^1_{g}|^2 + |e^2_{g}|^2\right) + \frac{n_{e,2}}{2}\left(|e^3_{g}|^2 + |e^4_{g}|^2\right).
\end{equation}
We then fit $\rho_{\mathrm{calc}}(r_0, \theta, \phi)$ to $\rho_{\mathrm{exp}}(r_0, \theta, \phi)$ with the free parameters $s$, $c_1$, $n_{a}$, $n_{e,1}$, and $n_{e,2}$ by minimizing the reliability factor $R_\mathrm{fit}$, defined as
\begin{equation}
\label{eq:R_fit}
R_\mathrm{fit} = \frac{\displaystyle \sum_{\theta,\phi}
\left|\rho_{\mathrm{exp}}(r_0, \theta,\phi)
- s\,\rho_{\mathrm{cal}}(r_0, \theta,\phi) \right|}
{\displaystyle \sum_{\theta,\phi}
\rho_{\mathrm{exp}}(r_0, \theta,\phi)}.
\end{equation}
Here, $s$ and $c_1$ represent a global scaling factor applied to $\rho_{\mathrm{calc}}$ and the mixing strength of $e_g$-$e_g'$ orbitals, respectively. In Fig.~\ref{main:Fig3}\textbf{g},\textbf{h}, the fit reproduces the data well, with $R_\mathrm{fit}=0.3~\%$.

\bigskip
\textbf{Unpolarized neutron diffraction}\\
Unpolarized neutron diffraction measurements were performed at the beamline D$9$ at Institut Laue-Langevin at $300\,$K. The neutron wavelength of $0.842\,$\AA\ was used. An as-grown single crystal with a mass of $183\,$mg was used. 
The crystal structure was refined using $134$ independent reflections, and the resulting extinction parameters and anisotropic displacement parameters were used in the multipole refinement of the polarized-neutron diffraction data. Averaging of symmetry-equivalent reflections and refinement of the crystal and magnetic structure were performed with \textsc{Mag2Pol}~\cite{Qureshi2019-tx}. Consistent with previous reports~\cite{Snow1952-aj,Takei1963-lc,Singh2025-wp}, the data show good agreement with the uniaxial collinear magnetic order (see Supplementary Information). While non-collinear altermagnetism is possible within a broader definition~\cite{Xiao2024-nd,Cheong2024-lt}, our neutron-diffraction data provide no experimental indication that a more complex magnetic structure is realized in CrSb.

\bigskip
\textbf{Polarized neutron diffraction}\\
Polarized neutron diffraction measurements were carried out at the spin-polarized hot neutron diffractometer D$3$ at the Institut Laue-Langevin at $300\,$K. The same single crystal as in the unpolarized neutron diffraction experiments was used, and the neutron wavelength was set to $0.829\,$\AA. The crystal was oriented with the $a^*$ and $c$ axes in the horizontal plane. Out-of-plane reflections were also measured thanks to a lifting-arm detector in the normal-beam configuration. A minimum magnetic field of $0.5\,$T was used to keep the polarization of the neutrons. Given fixed atomic positions in CrSb with the spin sequence in the order of Cr$_1$, Cr$_2$, as defined in Fig.~\ref{main:Fig2}\textbf{a}, two altermagnetic domains, $\uparrow\downarrow$ and $\downarrow\uparrow$, can be distinguished through the interference term in the flipping ratios ($\mathrm{FR}$); the two domains yield opposite signs of the magnetic form factors (see Supplementary Information). From the multipole refinement described below, the single crystal exhibited a naturally complete altermagnetic domain imbalance, where the $\downarrow\uparrow$ state is predominant within experimental uncertainty. This is indicated by large values of $\mathrm{FR}$; $\mathrm{FR}$ is always $1$ when two altermagnetic domains are equally populated (see Supplementary Information). More specifically, focusing on the strongest observed flipping ratio at the $0\,\overline{1}\,\overline{1}$ reflection, we observe $\mathrm{FR_{obs}}=2.43$. For domain ratios, $\downarrow\uparrow : \uparrow\downarrow = 1:0$, $0.75:0.25$, $0.5: 0.5$. $0.25: 0.75$, we expect $\mathrm{FR_{cal}}=2.38,\ 1.51,\ 1.0,\ 0.66$, respectively. This qualitative comparison for a single reflection also supports the naturally complete domain imbalance, inferred from the refinement.

The single domain formation can be understood in terms of the Ising nature of uniaxial altermagnets such as CrSb, where the N\'eel vector serves as the order parameter. In three dimensions, the Ising model undergoes a second-order phase transition at the N\'eel temperature ($T_{\mathrm{N}}$), entering a symmetry-broken phase that energetically favors a single-domain state. Such a large domain was also observed in the uniaxial antiferromagnet Cr$_2$O$_3$~\cite{Hayashida2022-id}. On the other hand, easy-plane altermagnets, such as MnTe and $\alpha$-Fe$_2$O$_3$, are characterized by the six-state Potts model due to the hexagonal symmetry of the lattice. This model exhibits a first-order phase transition at $T_\mathrm{N}$, leading to phase coexistence, as directly imaged in MnTe~\cite{Amin2024}.

Multipole refinements were performed with \textsc{Mag2Pol}~\cite{Qureshi2019-tx}. The azimuth angle was set to $\phi=90^\circ$ so that the axis definition matches with our notation. Fixing $n_l$ at $4$ and using $Z_l = 6.5$ as the initial parameter~\cite{Clementi1974-os} for Slater coefficients (see Supplementary Information), we first refine the isotropic model ($C_{00}$ and $Z_0$) using all the data. Here, $C_{lm}$ corresponds to the coefficient of $Q_{lm}$. This leads to $C_{00}=1.97(6),\, Z_0=6.8(1)$ with a strong anticorrelation $-0.85$ between them. Thus, we fix $Z_0 = 6.5$ to avoid overfitting. The fitting result of the isotropic model is shown in Supplementary Information. The level of agreement between the model and the data is quantified by $R_{\mathrm{FR}} =
\sum_i \left| \mathrm{FR}_{\mathrm{obs},i} - \mathrm{FR}_{\mathrm{cal},i} \right|
/ \sum_i \mathrm{FR}_{\mathrm{obs},i}$, where the index $i$ runs over all data points. The isotropic model results in $R_{\mathrm{FR}}=3.48\,\%$. We then fix $Z_l=Z_0=6.5$ for all $l$ because all the relevant multipoles arise from chromium $3d$ orbitals. Subsequently, we refine $C_{00}$, $C_{20}$, $C_{40}$, and $C_{43}$, obtaining $C_{00}=2.02(4),\, C_{20} = -0.05(3),\, C_{40}=0.13(8),\, C_{43}=0.21(6)$. The inclusion of $C_{43}$, which uniquely affects $l=\,$even, leads to good agreement between the calculated and observed flipping ratios for $l=\mathrm{even}$ as in Supplementary Information and as quantified by the improvement of $R_{\mathrm{FR}}$ to $2.13\,\%$. Using the multipole expansion, we fit the SD from theoretical calculations at a fixed radius $r=0.37\,$\AA, leading to $C_{43}=0.25$, in good agreement with the experiment. Here, we reverse the sign of $C_{43}$ for the theoretical SD,  since we take the other time-reversal partnered altermagnetic domain ($\uparrow\downarrow$), as compared to the experiment ($\downarrow\uparrow$). The existence of ferroic magnetic multipoles, arising from the concurrent antiferroic ordering of electric multipoles and magnetic dipoles, distinguishes CrSb from non-altermagnets with the same crystal structure.

\bigskip
\textbf{Electronic structure calculations}\\
We performed first-principles calculations for CrSb using Quantum ESPRESSO~\cite{Giannozzi2017-ag}, and the valence electron density (VED) and spin density (SD) are computed using the Julia package DiracBilinears.jl~\cite{Miki2025-us}. The crystal structure parameters of CrSb are taken from the experimental results, with lattice constants $a=4.123$~\AA\ and $c=5.47$~\AA~\cite{Kjekshus1969}. In the calculations, we use the Perdew-Burke-Ernzerhof (PBE) exchange-correlation functional~\cite{Perdew1996-ih} and optimized norm-conserving Vanderbilt pseudopotentials~\cite{Hamann2013-ot} provided in PseudoDojo~\cite{van-Setten2018-cx}. The cutoff energies for the wave function and charge density are set to $90$ and $360$~Ry, respectively, with an $8\times 8\times 8$ $\bm{k}$-point mesh. We found that spin-orbit coupling has little impact on the VED and that the magnetic order slightly modifies the VED (see Supplementary Information)

\bigskip

\textbf{Equivalence of VED and spin density (SD) in CrSb}\\
It has been predicted that the SD distribution shows an alternating arrangement on two magnetic sublattices in a wide class of altermagnets~\cite{Smejkal2022}. Like electric multipoles generate an anisotropic VED, magnetic multipoles are responsible for an anisotropic SD, and ferroically ordered magnetic multipoles serve as an order parameter of altermagnets~\cite{Bhowal2024,McClarty2024,Verbeek2024,Fernandes2024}. Since antiferroic electric multipoles combined with antiferroic magnetic dipoles directly generate ferroic magnetic multipoles, VED and SD are nearly equivalent in CrSb and related altermagnets based on $3d$ transition metals, as discussed below.

As confirmed in Fig.~\ref{main:Fig3}\textbf{e}-\textbf{h}, the VED at the Cr sites is well described by the high-spin state without double occupancy in any individual $d$ orbital. This means $\rho$ is equivalent to the spin-up VED $\rho_{\uparrow}$ or the spin-down VED $\rho_{\downarrow}$.
The SD $\rho_\mathrm{s}$ at sublattice $\alpha$ ($\alpha=\pm1$ for Cr$_1$, Cr$_2$) is therefore
\begin{equation}
    \rho_\mathrm{s}(\alpha) = \rho_{\uparrow}(\alpha) - \rho_{\downarrow}(\alpha) \approx \alpha\rho(\alpha),
\end{equation}
where $\rho_{\uparrow}\ (\rho_{\downarrow})$ is the VED of spin up (down), and $\rho(\alpha)= \rho_{\uparrow}(\alpha) + \rho_{\downarrow}(\alpha)$ is the total VED. Note that in this convention, spin density has the unit of charge per volume. Consistent with this relationship, the calculated SD shows a nearly equivalent pattern as compared to the experimental and calculated VED distributions (see Supplementary Information). Furthermore, the presence of the SOC does not qualitatively affect this near equivalence (see Supplementary Information).

The equivalence between VED and SD extends to other altermagnets provided two physical conditions are met.
(i) Hund’s coupling should exceed the crystal-field splitting of the relevant orbitals. Physically, this means that electrons align their spins parallel before occupying orbitals with electrons of opposite spin. (ii) The $d$ shell should be half-filled or less than half-filled. Condition~(i) also excludes cases in which the crystal field isolates a completely filled subset of $d$ orbitals. A representative example is a fully occupied $t_{2g}$ manifold with partially filled $e_g$ states, where the argument no longer applies. Notably, many high-temperature altermagnet candidates are based on $3d$ transition metals, for which Hund’s coupling typically dominates over crystal-field splitting, and thus condition~(i) is naturally satisfied.

Regarding condition~(ii), even when the $d$ manifold is more than half-filled, electron-hole symmetry implies that the VED and SD retain the same anisotropy pattern, although their magnitudes differ. Consider a case where the spin-up states are fully occupied, whereas the spin-down states are partially occupied. Writing the isotropic VED of the filled $d^5$ configuration as $\rho_0$ and the VED associated with an unoccupied (hole) state as $\rho_1$, we have $\rho_{\uparrow}=\rho_0$ and $\rho_{\downarrow}=\rho_0-\rho_1$. Hence the total VED is $\rho=2\rho_0-\rho_1$, whereas the SD is $\rho_{\mathrm{s}}=\rho_1$. The two thus share the same anisotropic component, differing only by an isotropic offset. We conclude that condition~(i) is the essential requirement.

We list prototypical altermagnets and show their average number of $d$ electrons per transition-metal site (see Supplementary Information). Altermagnets based on early transition metals, V and Cr, and one of the Mn-based materials host less-than-half-filled $d$ orbitals. Most Mn-based compounds exhibit a $d^5$ state, suggesting isotropic on-site VED and SD in the ionic limit. In summary, the VED--SD equivalence is expected to apply broadly to $3d$ transition-metal compounds, which constitute many of the most extensively studied altermagnets, including those with high transition temperatures. To induce a large on-site anisotropy and a concomitant large sublattice and spin splitting of electronic bands in momentum space, a well-defined ionic character is also helpful, yet even a metallic system can show a moderately large anisotropy, as demonstrated in the present work.

\bigskip
\textbf{Symmetry-adapted closest Wannier modeling}\\
Following the closest Wannier (CW) method~\cite{Ozaki2024-oz}, the CW tight-binding Hamiltonian $H^{\rm CW}$ was constructed by using the Python library SymClosestWannier~\cite{Oiwa2025-fz}.
We employed the pseudo-atomic orbitals (PAOs) from the pseudopotentials~\cite{Agapito2016-cm}, namely 3$s$, 3$p$, and 3$d$ PAOs for each Cr atom and 5$s$, 5$p$, and 4$d$ PAOs for each Sb atom, as the initial guesses for the CW functions (CWFs).
Since the symmetry properties of the CWFs $\{w_{n}^{\rm CW}(\bm{r})\}$ are common to those of the original PAOs, the symmetry-adapted multipole basis (SAMB) can be defined as the complete orthonormal matrix basis set $\left\{\mathcal{Z}_{j}\right\}$ in the Hilbert space of the CWFs, $\mathrm{Tr}(\mathcal{Z}_{i} \mathcal{Z}_{j}) = \delta_{ij}$.
The CW tight-binding Hamiltonian $H^{\rm CW}$ can thus be expressed as a linear combination of $\mathcal{Z}_{j}$, $\mathcal{H}^{\rm SymCW} = \sum_{j} z_{j} \mathcal{Z}_{j}$, with coefficients given by $z_{j} = \mathrm{Tr}(\mathcal{Z}_{j} H^{\rm CW})$. This accurately reproduces the calculated band structure (see Supplementary Information). Using the SAMBs associated with the electric crystal potentials $\hat{Q}_i(\alpha)\ (i=lm)$, the VED around Cr sublattice $\alpha=\pm 1$, $\rho(\alpha)$, from the CW method can be approximately expanded as a linear combination of the $\hat{Q}_i(\alpha)$:
\begin{equation}
\label{eq:ved_cw_main}
\rho(\alpha) \simeq \sum_{i} \langle \hat{Q}_{i}(\alpha) \rangle \sum_{mn} [\hat{Q}_{i}(\alpha)]_{mn} w_{m}^{\rm CW}(\bm{r}) w_{n}^{\rm CW *}(\bm{r}).
\end{equation}

\bigskip
\textbf{Minimal models for \textit{g}-wave altermagnetism}\\
Here, we discuss two extreme cases for altermagnetism: $(1)$ isotropic VED with $g$-wave hopping and $(2)$ $g$-wave VED without $g$-wave anisotropy in hopping. See Supplementary Information for detailed discussions.

Case~$(1)$ corresponds to the most widely studied toy model of an altermagnet. As discussed in Ref.~\cite{Roig2024}, a minimal Hamiltonian of $g$-wave altermagnetism  
reads

\begin{equation}
    H(\bm{k}) = h(\bm{k})\tau_0 + t_{z}(\bm{k})\tau_z + J_z\tau_z\sigma_z,
\end{equation}
where $h(\bm{k}) \tau_{0}\sigma_0$ ($\tau_0,\, \sigma_0$: identity matrix) and $\tau_{z} (\sigma_z)$ 
denote the sublattice-independent dispersion and the Pauli matrix that describes the sublattice (spin) degrees of freedom, respectively. This model considers only the magnetic sites on the lattice and incorporates effects of ligands through the anisotropic hopping amplitude $t_{z}(\bm{k})$. The form of $t_{z}(\bm{k})$ was obtained from a symmetry argument as $t_{z}(\bm{k}) = tf_y(f_y^2-3f_x^2)\sin k_z$ with $f_x = \sin k_x + \sin(k_x/2) \cos (\sqrt{3}k_y/2)$ and $f_y = \sqrt{3} \cos(k_x/2) \sin (\sqrt{3}k_y/2)$~\cite{Roig2024}. We expand $t_z(\bm{k})$ and find that this model involves long-range $8$th and $16$th nearest-neighbor (NN) hopping between magnetic atoms (see Supplementary Information). In CrSb, the distance of the $8$th NN is $9\,\text{\AA}$. The lowest-order contribution to the spin splitting reads
\begin{equation}
    g_2(\bm{k}) \propto J_ztk_zk_y(3k_x^2-k_y^2),
\end{equation}
where the subscript $2$ of $g_2(\bm{k})$ denotes the perturbative order in the microscopic couplings (here, $J_z$ and $t$).

A more realistic model for case~$(1)$ is given by the $1$st NN hopping between magnetic and ligand atoms (see Supplementary Information). In this model, the $g$-wave symmetry is embedded in intersite $e_g$-$e'_g$ mixing; the hopping matrix is the same as $\hat{Q}_{43}$. This model gives rise to the $g$-wave spin splitting (see Supplementary Information), whose lowest-order contribution is given by
\begin{equation}
    g_7(\bm{k}) \propto J_zt_1^2t_2^4k_zk_y(3k_x^2-k_y^2),
\end{equation}
where $t_1$ and $t_2$ are components of the $1$st NN anisotropic hopping $t'$ without and with $g$-wave anisotropy, respectively.

Regarding the other extreme case~$(2)$, we find that a minimal model involves $3$rd NN hopping between magnetic atoms (see Supplementary Information). The model does not explicitly include ligand atoms; their effects are instead incorporated as anisotropy of the VED that results from $Q_{43}$. Essentially, a combination of the anisotropic VED and hopping $t$ without $g$-wave anisotropy gives rise to $g$-wave spin splitting (see Supplementary Information), and the lowest-order contribution is calculated as
\begin{equation}
    g_4(\bm{k}) \propto J_zt^2\phi k_zk_y(3k_x^2-k_y^2),
\end{equation}
where $\phi = q_{43}/2\sqrt{2}$, and $q_{43}$ is the coefficient of $\hat{Q}_{43}$ in $\mathcal{H}$ [Eq.~\eqref{eq:CEF_main}]. In general, realistic models with short-range hopping require a larger number of degrees of freedom and higher-order contributions to the spin splitting.

\bigskip

\newpage

\clearpage
\textbf{Acknowledgments}\\
We are grateful to N. Spaldin, M.~T. Birch, O. Zaharko and Y. Motome for helpful discussions. The authors are indebted to S.~Aoyagi and N.~Kanazawa, who wrote software that accelerated the compression of large VED datasets.
This work was supported by JSPS KAKENHI Grant Nos. JP22K20348, JP23K13057, JP23H04869, JP24H01607, JP24H01604, JP25K17336, JP24K17006, JP24H01644, JP23KJ0298, JP25K24552, JP26H00644 and JP26H01290 as well as JST CREST Grant Nos. JPMJCR1874 and JPMJCR20T1 (Japan) and JST PRESTO Grant No. JPMJPR259A and JST FOREST Grant No. JPMJFR2238, and No. JPMJFR2362 (Japan). It was also supported by JST as part of Adopting Sustainable Partnerships for Innovative Research Ecosystem (ASPIRE), Grant Number JPMJAP2426. M.H. is supported by the Deutsche Forschungsgemeinschaft (DFG, German Research Foundation) via Transregio TRR 360 – 492547816.
J.~M.~acknowledges funding from the DFG under Project No. 547968854. The synchrotron single-crystal X-ray experiments were performed at BL02B1 in SPring-8 with the approval of RIKEN (Proposal No. 2025B1602, 2025B1714 and 2026A1920). The neutron diffraction experiments were performed at the D$3$ and D$9$ beamlines of Institut Laue-Langevin (Grenoble, France). We acknowledge the Polish Ministry of Education and Science's decision Nr.~2023/WK/08 to fund the scientific membership of Poland at the ILL, which made the neutron experiment possible.

\end{document}